\documentclass[aps,prl,twocolumn,groupedaddress]{revtex4}

\usepackage{graphicx}
\usepackage[draft]{hyperref}
\usepackage{amsmath}

\begin{document}


\title{Magnetic metamaterials in the blue range using aluminum nanostructures}

\author{Yogesh Jeyaram,$^{1}$ Shankar K. Jha,$^{1}$
Mario Agio,$^2$ J\"org F. L\"offler,$^{1}$ Yasin Ekinci,$^{1,3,*}$}

\affiliation{
$^1$Laboratory of Metal Physics and Technology, Department of Materials,
ETH Zurich, 8093 Zurich, Switzerland\\
$^2$Nano Optics Group, Laboratory of Physical Chemistry,
ETH Zurich, 8093 Zurich, Switzerland\\
$^3$Paul Scherrer Institute, 5232 Villigen-PSI, Switzerland\\
$^*$Corresponding author: yasin.ekinci@mat.ethz.ch
}

\begin{abstract}
We report an experimental and theoretical study of the optical
properties of two-dimensional arrays of aluminum nanoparticle
in-tandem pairs. Plasmon resonances and effective optical constants
of these structures are investigated and strong magnetic response as
well as negative permeability are observed down to 400 nm
wavelength. Theoretical calculations based on the
finite-difference time-domain method are performed for
various particle dimensions and lattice parameters, and are found to
be in good agreement with the experimental results.  The results
show that metamaterials operating across the whole visible
wavelength range are feasible.
\end{abstract}


\maketitle

In the last decade, a new class of optical materials known as
metamaterials have emerged and attracted significant interest. These
are metallic structures engineered at the subwavelength scale to
exhibit novel optical properties such as negative refractive index
or magnetic activity at high frequencies~\cite{shalaev07}. After the
first experimental demonstration in the microwave regime, the
operation frequency of metamaterials has experienced a tremendous
progress within a decade due to novel designs and nanofabrication
techniques~\cite{soukoulis07}. In particular, metal/dielectric/metal
multilayers, such as cut-wire nanopairs, fishnet nanostructures, and
nanoparticle pairs, have enabled metamaterials operating at optical
frequencies. These structures exhibit strong antisymmetric
eigenmodes providing magnetic resonances and negative permeability,
which is the prerequisite of negative refractive index metamaterials
~\cite{huang06,pakizeh06,cai07,pakizeh08,ekinci08}. With such
designs magnetic metamaterials at about 1.2 $\mu$m have been
realized using interference lithography~\cite{feth07}. Record
wavelengths of metamaterials have been achieved by Chettiar \emph{et
al.} demonstrating negative refractive index at 813 nm and negative
permeability at 772 nm~\cite{chettiar07} and by Yuan \emph{et al.}
demonstrating negative permeability at 725 nm~\cite{yuan07}.
Moreover, 1D-metamaterials have been demonstrated exhibiting a
magnetic resonance~\cite{xiao09} and a negative
permeability~\cite{schweizer07} at about 630 nm.

In these aforementioned and many other works demonstrating
metamaterials with negative  refractive index or negative
permeability mostly Au and Ag were used, which are, in fact, the
conventional choice in plasmonics. On the other hand, Al is also a good
optical material because of its low absorption and large real part of
the dielectric constant. Whereas Au and Ag exhibit interband
absorptions below the wavelengths of about 590 nm and 350 nm,
respectively, which limit their applications in optics and
plasmonics, Al has low absorption down to 200 nm due to its
free-electron-like character and high bulk plasmon
frequency~\cite{palik98}. These properties make Al an ideal
candidate for plasmonic applications at short
wavelengths~\cite{ekinci07,lakowicz07,ekinci08b, mohammadi09}.

In this letter we study the plasmonic properties of arrays of Al
in-tandem particle pairs. Ordered 2D arrays of Al/Al$_{2}$O$_{3}$/Al
nanoparticles show strong plasmonic resonances of hybridized
eigenmodes due to the near-field coupling between nanoparticles. We
demonstrate that it is possible to tune the magnetic resonances down
to 400 nm and obtain even negative permeability covering the whole
visible range.

\begin{figure}[htb]
\centerline{\includegraphics[width=8cm]{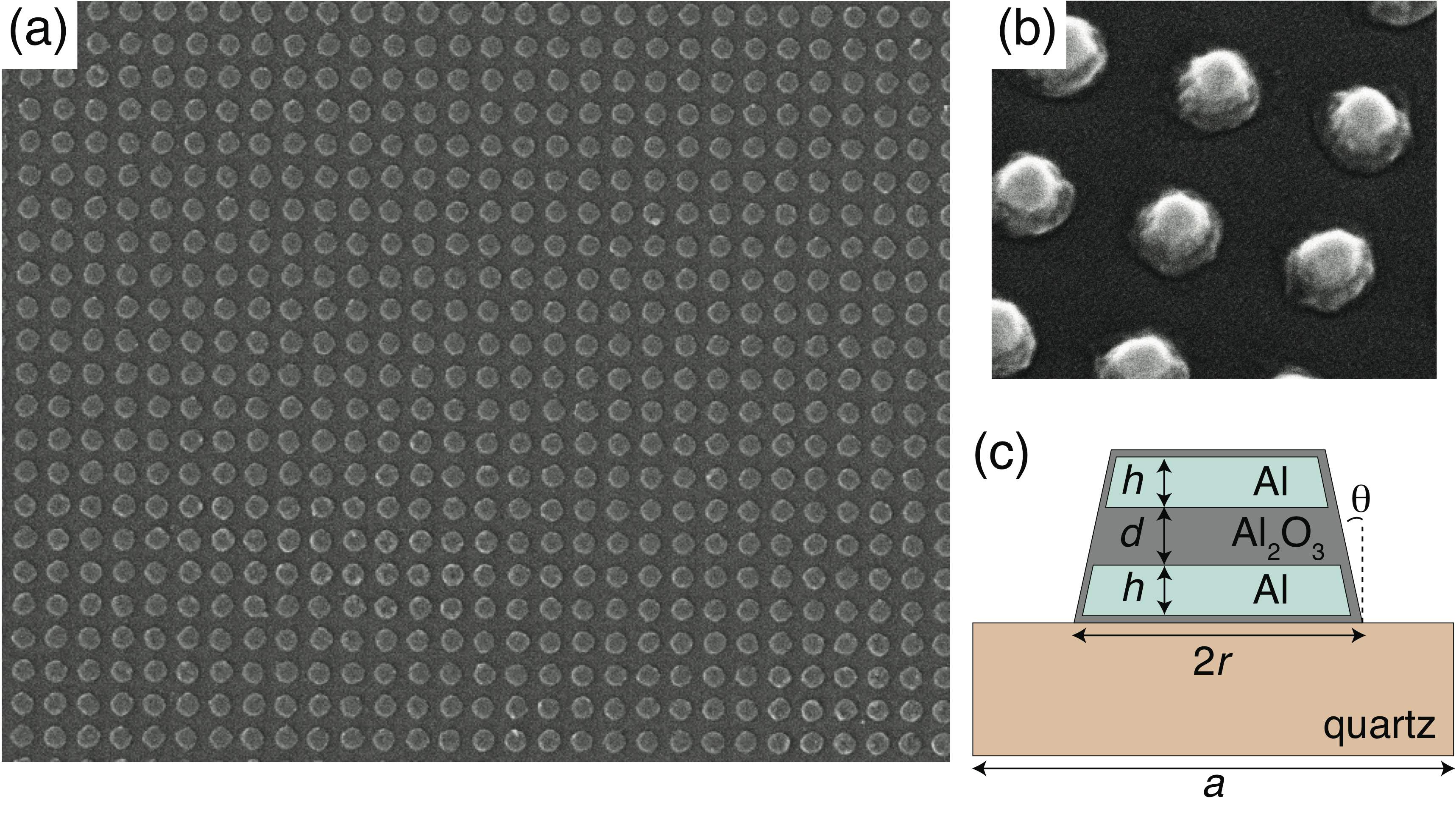}}
\caption{\label{figure1}(a) SEM image of one of the fabricated samples (diameter, $2r$ = 120 nm and period $a=$ 200 nm). (b) SEM
image at an oblique angle with details on the vertical profile of
one sample ($2r$ = 105 nm). (c) Model used for the FDTD calculations
($h$ = 21 nm, $d$ = 24 nm, $a$ = 200 nm, $\theta=20^o$,
$n_\mathrm{Al_2O_3}=1.77$, $n_\mathrm{quartz}=1.46$, $n_\mathrm{Al}$
is from~\cite{palik98} and $2r=135, 120, 105, 87, 82, 72$ nm; a
3-nm-thick Al$_2$O$_3$ coating layer is also taken into account). }
\end{figure}

Two-dimensional arrays of Al nanoparticles pairs on quartz
substrates (HPFS) were fabricated using extreme-ultraviolet
interference lithography (EUV-IL) and sequential deposition of Al,
Al$_{2}$O$_{3}$, and Al with a subsequent lift-off process. EUV-IL
provides high-resolution structures over large areas with perfect
periodicity~\cite{solak07}. Moreover, there is no need for a
conduction layer such as ITO which is used in e-beam lithography.
Figure~\ref{figure1}(a) shows a top-down scanning electron microscope
(SEM) image of a typical sample. Figure~\ref{figure1}(b) is another
SEM image taken at an oblique angle, demonstrating the sandwich-like
geometry of the fabricated in-tandem nanoparticle pairs, the
schematic cross-section of which is shown in Fig.~\ref{figure1}(c). The size
of the fabricated structures varied from $2r=$ 72 nm to 135 nm (in base
diameter) with a fixed period of $a=$ 200 nm. The
thicknesses of the Al and Al$_{2}$O$_{3}$ layers were 21 nm and 24
nm, respectively. The area of an array was 400 $\mu$m $\times$ 400
$\mu$m, simplifying optical characterization substantially.

\begin{figure}[htb]
\centerline{\includegraphics[width=8.3cm]{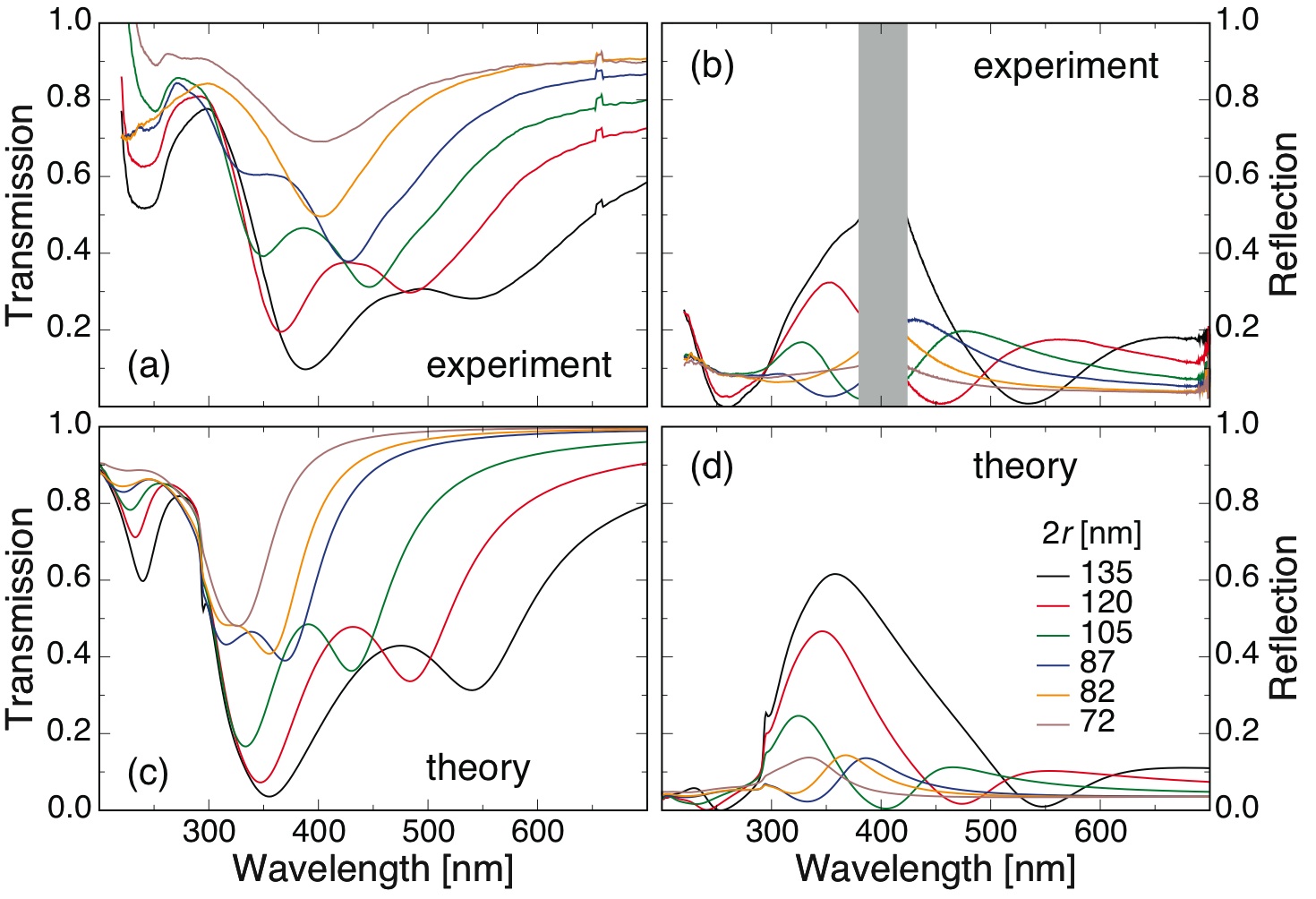}}
\caption{\label{figure2}Transmission ((a) and (b)) and reflection
((c) and (d)) spectra for samples with different disk diameters
$2r$. (a) and (c) experimental, (b) and (d) theoretical FDTD
results. Note that the reflection spectra in (b) were measured with
two different setups (200-400 nm and 400-700 nm). A small range of
the spectra (shaded region between 380-420 nm) is not shown because
the results were not reproducible because of the low signal-to-noise
ratios of both setups in this region. }
\end{figure}

Transmission spectra of the structures were measured with a
home-made micro-spectroscopy setup, which enables spectroscopy in
the visible and UV ranges. The
reflection spectra were measured with the same home-made setup in
the UV range and with a standard optical microscope in the visible
range. Since the structures are isotropic on the substrate plane, no
special polarization optics was used. Figures~\ref{figure2}(a) and
(b) show the transmission and reflection spectra of the samples with
different diameters. The transmission spectra exhibit two dips
associated with hybridized modes resulting from the dipolar coupling
between the individual nanoparticles of a pair. The resonance at
shorter wavelengths can be identified as a symmetric resonance or
electric dipolar resonance while the one at longer wavelengths is
attributed to the antisymmetric or magnetic dipolar
resonance~\cite{pakizeh06,ekinci08}. These two resonances are also
evident in the reflection spectra (see Fig.~\ref{figure2}(b)) where
the resonance peaks appear asymmetric due the Fano-type of
interference between the eigenmodes~\cite{ekinci08}.  The antisymmetric mode
arises due to the induced dipole vectors in the two metal layers
being opposite to each other. This creates a magnetic dipole moment
which couples to the magnetic component of the incident field and
thus gives rise to effective permeability diverging from unity. As
seen in Fig.~\ref{figure2}(b) the magnetic resonance can be easily
tuned by varying the particle diameter for constant metal and
dielectric thickness.

Finite-difference time-domain (FDTD) simulations were performed
using the schematic model shown in Fig.~\ref{figure1}(c), which
represents relatively simple but realistic structural parameters of
the fabricated structures. A natural Al$_{2}$O$_{3}$ layer of 3 nm
is assumed to cover the nanoparticles at the Al-air and Al-substrate
interfaces, which is known to be formed immediately when Al is
exposed to air or oxygen-rich substrate~\cite{gertsman05}.  For
dielectric constants of Al and Al$_{2}$O$_{3}$ we used the reported
values from the literature~\cite{palik98}. Periodic boundary
conditions were applied to account for the effects of array on
optical response. Figure~\ref{figure2} shows good agreement
between the experimental and the calculated spectra, especially for
the samples with larger nanoparticles. The small variation between
the theoretical and experimental results are due to discrepancies
between the exact geometry and optical constants between the modeled
and fabricated structures. Moreover, the size and the shape
dispersion of the samples resulting from the finite grain size of deposited
Al could not be taken into account in the theory.

\begin{figure}[htb]
\centerline{\includegraphics[width=8.3cm]{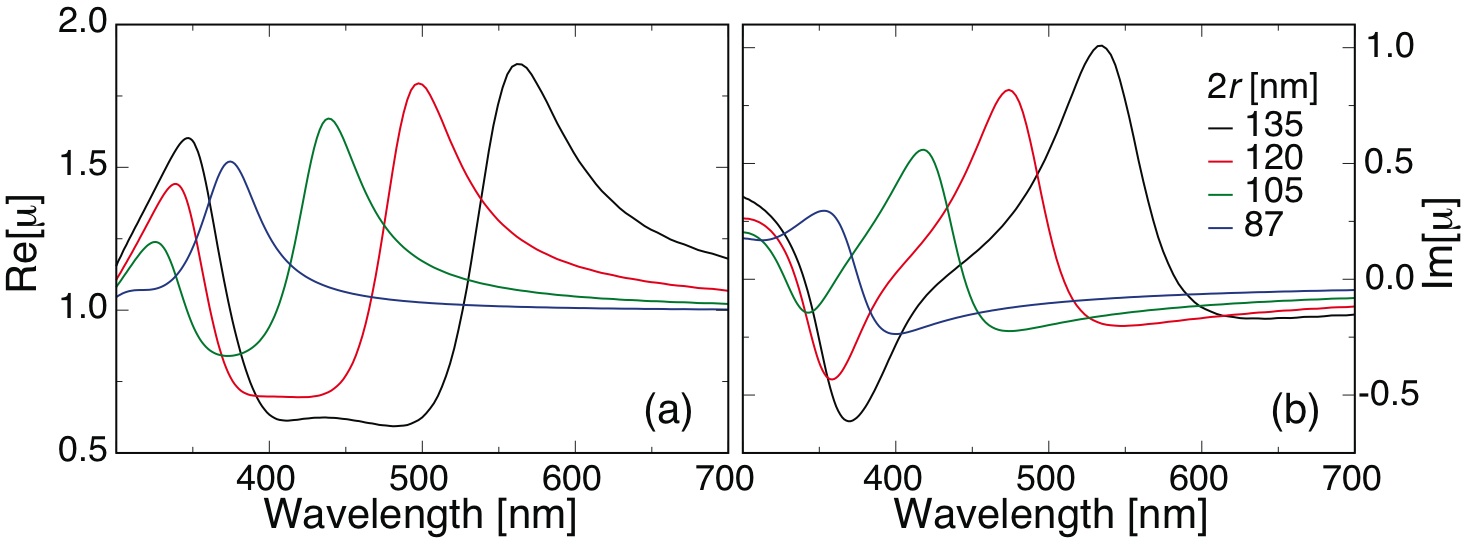}}
\caption{\label{figure3}(a) Real and (b) imaginary part of the
relative magnetic permeability $\mu$ extracted from FDTD simulations
for the fabricated samples.}
\end{figure}
Since the present structures exhibit significant magnetic resonances
we can expect a strong modulation of the effective magnetic
permeability. We extracted the effective optical constants from the
complex reflectance and transmittance
coefficients~\cite{smith02,shalaev05}. Figure~\ref{figure3} displays
the complex relative magnetic permeability $\mu$ for selected
samples. Different nanodisk radii $r$ provide strong magnetic
resonances covering the  visible and the UV spectral ranges with
strong modulations of Re[$\mu$] between 0.5 and 2 and without a
significant change in the figure of merit $|$Re[$\mu$]/Im[$\mu$]$|$.
The results show how the amplitude and the wavelength of the
effective permeability of the present metamaterial structures can be
easily tuned.

\begin{figure}[htb]
\centerline{\includegraphics[width=8.3cm]{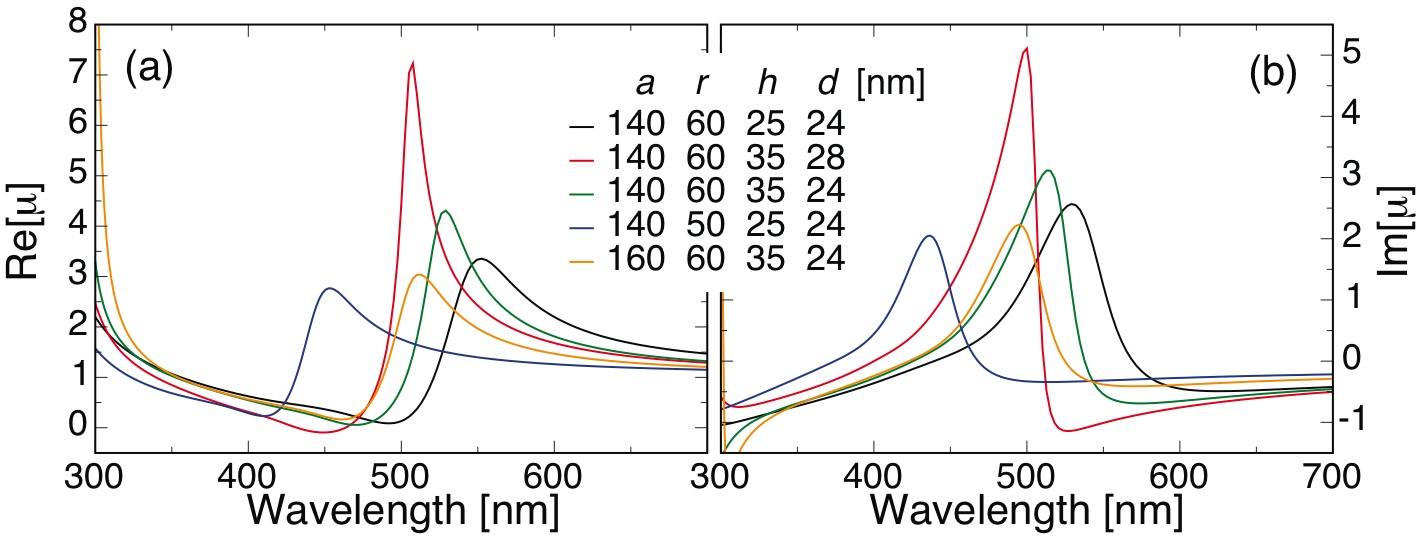}}
\caption{\label{figure4}(a) Real and (b) imaginary part of the
relative magnetic permeability $\mu$ extracted from FDTD simulations
for the structural parameters provided in the legend. The oxide
coating has been neglected.}
\end{figure}

Having demonstrated the tunability of the magnetic resonance
wavelength, we performed extensive FDTD simulations to optimize the
structural parameters in order to study the tunability of the
amplitude of the magnetic resonance. Particular interest was paid on achieving
negative permeability at short wavelengths within the structural
parameters feasible with standard lithographic processes.
The extracted $\mu$ for some representative cases is shown in
Fig.~\ref{figure4}. A strong magnetic response is obtained at around
500 nm, with $\mu$ as large as 7. A negative permeability at about
450 nm is achieved, notwithstanding its relatively low figure of
merit. As seen in the figure, in changing the lattice parameter from
160 nm to 140 nm, the magnetic resonance of the same nanoparticle
pair exhibits a red shift. Similar to electric dipole interactions
in nanoparticles arrays~\cite{lamprecht00}, we show the evidence of
magnetic dipole coupling, which deserves a detailed investigation in
future studies.

In summary, we have demonstrated that magnetic metamaterials can be
obtained down to the blue  wavelength range with 2D arrays of Al
in-tandem nanoparticles. Thanks to the optical properties of Al,
negative permeability is shown to be feasible paving the way towards
negative refractive-index metamaterials in the whole visible range.
Even though the figure of merit is relatively low, the promise of
the present work should invoke further studies with new designs like
nanohole array or fishnet structures~\cite{dolling07}, which may
provide negative permeability with a higher figure of merit and even
negative refractive index.  The Al metamaterials studied in
this work also offer new opportunities in other applications of
metamaterials, such as in impedance matching to maximize power transfer
from vacuum into material\cite{lee05}, magnetic waveguides as
building blocks of optical circuitry\cite{liu09}, novel optical
materials supporting multiple resonances \cite{cai07}, and as
unidirectional optical nanoantennas for surface-enhanced
spectroscopy~\cite{pakizeh09}.

This work was supported by ETH Research Grant TH-29/07-3.
MA thanks Vahid Sandoghdar for continuous support and encouragement.
Part of this work was performed at the Swiss Light Source (SLS),
Paul Scherrer Institute, Switzerland.

\end{document}